\documentclass[prd,aps,twocolumn]{revtex4}
\usepackage{graphicx,url,color}
\usepackage{amsmath,amssymb}

\def\be{\begin{equation}}
\def\ee{\end{equation}}
\def\ba{\begin{align}}
\def\ea{\end{align}}

\def\lsim{\raise0.3ex\hbox{$\;<$\kern-0.75em\raise-1.1ex\hbox{$\sim\;$}}}
\def\gsim{\raise0.3ex\hbox{$\;>$\kern-0.75em\raise-1.1ex\hbox{$\sim\;$}}}

\def\theta{\vartheta}

\def\Xmax{$X_{\max}$}
\def\X2{RMS($X_{\max})$}

\begin{document}

\title{A minimal model for extragalactic cosmic rays and neutrinos}

\author{M.~Kachelrie\ss$^{1}$}
\author{O.~Kalashev$^{2}$}
\author{S.~Ostapchenko$^{3,4}$}
\author{D.~V.~Semikoz$^{5,6}$}

\affiliation{$^{1}$Institutt for fysikk, NTNU, Trondheim, Norway}
\affiliation{$^{2}$Institute for Nuclear Research of the Russian Academy of Sciences, Moscow, Russia}
\affiliation{$^{3}$Frankfurt Institute of Advanced Studies, Frankfurt, Germany}
\affiliation{$^{4}$Skobeltsyn Institute of Nuclear Physics, Moscow State University, Moscow, Russia}
\affiliation{$^{5}$AstroParticle and Cosmology (APC), Paris, France}
\affiliation{$^{6}$National Research Nuclear University (MEPHI), Moscow, Russia}

\begin{abstract}
We aim to explain in a unified way the experimental data on ultrahigh
energy cosmic rays (UHECR) and neutrinos, using a single source class
and obeying limits on the extragalactic diffuse gamma-ray background (EGRB). 
If UHECRs only 
interact hadronically with gas around their sources, the resulting diffuse 
CR flux can be matched well to the observed one, providing at the same 
time large neutrino fluxes. 
Since the required fraction of heavy nuclei is, however, rather large, 
air showers in the Earth's atmosphere induced by UHECRs with energies 
$E\gsim 3\times 10^{18}$\,eV would reach in such a case their maxima too high.
Therefore additional 
photo-hadronic interactions of UHECRs close to the accelerator 
have to be present, in order to modify the nuclear composition of 
CRs in a relatively narrow energy interval. We include thus both photon and 
gas backgrounds, and combine the resulting CR spectra with 
the high-energy part of the Galactic CR fluxes predicted by the escape 
model. As result, we find a good description of experimental data on
the total CR flux,  the mean shower maximum depth \Xmax\ and
its width \X2 in the whole energy range above $E\simeq 10^{17}$\,eV. 
The predicted high-energy neutrino flux matches IceCube measurements, 
while the contribution to the EGRB is of order 30\%.
\end{abstract}


\pacs{xx}

\maketitle

\section{Introduction}

A major motivation for the construction of km$^3$ neutrino telescopes has 
been the goal to identify the sources of ultrahigh energy cosmic rays (UHECR):
While deflections of charged CRs in magnetic fields have prevented so far
a successful correlation of their arrival directions with potential sources
even at the highest energies,  photons and neutrinos point back to 
their sources. These neutral secondaries are produced by UHECRs
interacting with gas or photons in their sources, and with cosmic microwave
and other background photons during propagation. Any process involving 
hadronization leads mainly to the production of pions, and isospin symmetry
fixes then the ratio of charged to neutral pions produced.
The production of neutrinos is thus intimately tied to the one of 
photons, and both depend in turn on the flux of primary cosmic rays.
This is the basis of the multi-messenger approach to UHECR 
physics~\cite{Gaisser:1994yf}.

In this work, we will address the question if a single source class
can explain $i)$ the extragalactic CR flux, $ii)$ its nuclear 
composition and $iii)$ the observed neutrino flux in IceCube. 
Moreover, we  will require that $iv)$ the accompanying photon flux 
is only a subdominant contribution to the extragalactic gamma-ray 
background (EGRB) measured by Fermi-LAT~\cite{EGRB}. Finally, 
$v)$ the model should be consistent with an early galactic to extragalactic 
transition. No model has been developed yet
which satisfies all five requirements. Since the neutrino flux measured 
by IceCube is high~\cite{nuflux,nuflux2}, i.e.\ close to the cascade 
limit~\cite{casc}, combining  $iii)$ 
and $iv)$ is challenging for many source classes. Moreover, existing 
models aiming to reproduce the observed nuclear composition of UHECRs fail
to produce sizable neutrino fluxes~\cite{others1,Unger:2015laa}. 
In contrast, models leading to large neutrino fluxes in the 0.1--1\,PeV 
energy range use typically proton primaries with 10--100\,PeV energies, 
without a direct connection to measurements of UHECR composition~\cite{others2,more}.

\section{Constraints}

Let us now explain these conditions in more detail. Additionally to the 
all-particle CR spectrum, data on the primary composition have become 
available in the last years: The Auger collaboration derived the fraction 
of four different 
elemental groups above $6\times 10^{17}$\,eV~\cite{Aab:2014aea},
while the KASCADE-Grande experiment measured the composition up to 
$2\times 10^{17}$\,eV~\cite{dataKG}. These measurements can be summarized
as follows:  First, the proton fraction amounts to
$\sim 40$--60\% in the energy range between $7\times 10^{17}$\,eV and 
$7\times 10^{18}$\,eV and decreases afterwards, while the fraction of 
intermediate nuclei increases. Second, the iron fraction in the 
energy range between $7\times 10^{17}$\,eV and $2\times 10^{19}$\,eV is 
limited by $\lsim $15--20\%. Despite both theoretical and 
experimental uncertainties, the following conclusions can be 
drawn: First, the Galactic contribution to the observed CR spectrum has to die 
out around $7\times 10^{17}$\,eV. This inference is supported by limits 
on the CR dipole anisotropy which require a transition below 
$\sim 10^{18}$\,eV in case of a light composition~\cite{dipole}.
Consequently, we demand an early Galactic-extragalactic transition and
have to explain therefore the ankle as a feature in the extragalactic
CR spectrum. Second, the composition measurements are inconsistent with a 
strong dominance of either protons or iron nuclei. We assume therefore
that a mixture of nuclei is injected in the source, with a rigidity
dependent maximal energy.

The main part of the EGRB is attributed to unresolved 
blazars~\cite{Neronov:2011kg,dimauro,Ajello:2015mfa,Fermiun}. This
makes blazars and in particular BL Lacs attractive neutrino sources, 
since their contribution to the EGRB is much larger than the one from 
other sources. An attempt to connect the observed UHECR proton flux 
with neutrino and gamma-ray data was performed in Ref.~\cite{escape3}.  
However, correlation studies of arrival directions of muon neutrinos 
with Fermi blazars showed that blazars cannot be the main source of 
IceCube neutrinos~\cite{bllac_fermi}. Thus leptonic models are favored 
to explain the main part of the photon flux from  blazars. 
As a result, neutrino sources should give a subdominant contribution to 
the EGRB.

\section{Source model}

To be specific, we will assume in the following that UHECRs are accelerated
by (a subclass of) active galactic nuclei (AGN). Acceleration could proceed 
either via shock acceleration in accretion shocks~\cite{KE86} or via 
acceleration in regular fields~\cite{Bl76,Lo76,BZ77,MT82,Neronov:2007mh} close 
to the supermassive black hole (SMBH). Alternatively, UHECRs could be 
accelerated in
large-scale radio jets~\cite{RB} or via a two-step acceleration process
in the jet and a radio-lobe \cite{lobes}. 
We neglect the details of the acceleration process, and assume that
the energy spectra of nuclei follow a power-law with a rigidity dependent 
cutoff
\be
j_{\rm inj}(E) \propto E^{-\alpha} \exp[-E/(ZE_{\max})] \,.
\ee
Subsequently, the CR nuclei 
diffuse first through a zone dominated
by photo-hadronic interactions, before they escape into a second zone 
dominated by hadronic interactions with gas. The propagation in both zones
is modeled as a one-dimensional process, determined by the ratio $\tau$
of the interaction rate $R_{\rm int}$ to the escape rate $R_{\rm esc}$.

The spectra of AGN show a characteristic blue bump,  produced by UV photons 
emitted from the accretion disk surrounding the SMBH~\cite{bluebump}. At lower 
energies, the spectra are dominated by thermal emission from dust surrounding 
the SMBH.
We assume that these IR photons provide the main interaction target, and
approximate their energy spectrum  by a thermal 
distribution. In the next section, we will see that interactions on
IR photons will result in a suppression of the heavier nuclei fluxes at
the desired energy range. We parametrise the interaction 
depth $\tau$ as $\tau_0^{p\gamma}=R_{\rm int}/R_{\rm esc}$, using protons 
with energy $E_0=10$\,EeV as reference.
We assume that the accelerated CR nuclei escape from the region filled
with thermal photons in a diffusive way, so that their escape rate 
$R_{\rm esc}$ is proportional to $(E/Z)^{\delta_{p\gamma}}$. For the numerical 
simulations, we use the  open source code~\cite{Kalashev:2014xna} which is 
based on kinetic equations in one dimension. 
Since the propagation time is the same for all particles in the code,  
diffusion of charged particles is taken into account by multiplying the 
interaction rates by the 
escape times and setting the propagation time equal to one.
Neutrons escape freely.

In the 2nd zone, we model both interactions with the gas and the escape 
of CRs as a Monte Carlo process, using 
QGSJET-II-04~\cite{qgsjetII} to describe nucleus-proton collisions.
We assume now that charged CRs diffuse in an extended halo and model escape 
and interactions in the leaky-box picture. The produced neutrons escape 
again freely. The source is  then fully described specifying the interaction 
depth $\tau_0^{pp}=R_{\rm int}/R_{\rm esc}$ of protons at the reference energy 
$E_0=10$\,EeV and the energy dependence of the escape rate,
$R_{\rm esc}=R_0 [E/(ZE_0)]^{\delta_{pp}}$.

The spectrum of particles exiting the source is then used in the third 
step as an  ``effective injection spectrum'', from which we calculate 
using again the code~\cite{Kalashev:2014xna} the 
resulting diffuse flux, taking into account the distribution 
of sources as well as the interaction of protons, electrons and photons
with the EBL and the CMB. We employ as our  baseline EBL 
model the one of Ref.~\cite{EBL}, but have verified that other models, e.g.\ 
the more recent one~\cite{Stecker:2016fsg}, lead only to small variations
in the CR flux and composition.

For the cosmological evolution of AGNs we use the parameterization 
obtained in Ref.~\cite{Hasinger:2005sb}, 
\be
\rho(z)=
 \begin{cases}
  (1+z)^m & \text{for } z<z_c ,
\\
 (1+z_c)^m
& \text{for } z_c<z<z_d ,
\\
 (1+z_c)^m 10^{k(z-z_d)}
& \text{for } z>z_d ,
 \end{cases}
\ee
choosing as parameters  $m=3.4$, $z_c=1.2$, $z_d=1.2$, and $k=-0.32$.
They correspond to the ones derived in Ref.~\cite{Hasinger:2005sb} for 
$\log(L_X/{\rm erg})=43.5$.

As second option, we use the parameterization for BL Lac/FR1 evolution 
presented in~\cite{dimauro}.
We describe the cosmological evolution of BL Lac/FR I sources as
\begin{equation}
 N_c(z) \propto 
 \int_{L_{\gamma}^{min}}^{L_{\gamma}^{max}} \rho(z,L_{\gamma}) L_{\gamma} dL_{\gamma} ,
\end{equation}
where $\rho(z,L_{\gamma})$ is the $\gamma$-ray luminosity function, i.e.\ the
number of sources per  comoving volume and luminosity. For $\rho(z,L_{\gamma})$ 
we adopt the Luminosity-Dependent Density Evolution (LDDE) model of 
Ref.~\cite{dimauro}  given by
\begin{eqnarray}
\label{eq:ldde1}
   \rho(z,L_{\gamma})= \rho(L_{\gamma}) \, e(z,L_{\gamma}) ,
\end{eqnarray}
with 
\begin{eqnarray}
\label{eq:ldde0}
   \rho(L_{\gamma})= 
	\frac{A}{\log{(10)}L_{\gamma}}\left[ \left( \frac{L_{\gamma}}{L_{c}} \right)^{\gamma_1} + \left( \frac{L_{\gamma}}{L_{c}} \right)^{\gamma_2} \right]^{-1} ,
\label{eq:ldde2}\\
   e(z,L_{\gamma}) = \left[ \left( \frac{1+z}{1+z_c(L_{\gamma})} \right)^{p_1} + \left( \frac{1+z}{1+z_c(L_{\gamma})} \right)^{p_2} \right]^{-1} ,
\end{eqnarray}
and
\begin{eqnarray}
\label{eq:ldde3}
  z_c(L_{\gamma}) = z^{\star}_c\, \left( \frac{L_{\gamma}}{10^{48} \rm{\,erg\,s}^{-1}}\right)^{\alpha} \,.
\end{eqnarray}
We use the numerical values given in Table 3 of Ref.~\cite{dimauro} for
the free parameters of this model.  
The evolution of the effective source density 
as function of redshift is shown in Fig.~3 of Ref.~\cite{unified}. This
figure illustrates that,  in contrast to average AGNs, 
the number density of BL Lac and FR I galaxies 
peaks at low redshift, $z\lesssim 1$. Thus their evolution is similar to that 
of  galaxy clusters. In fact, most of the FR I sources, which are the 
parent population of BL Lacs reside in the centres of the dominant central 
elliptical galaxies of galaxy clusters (cD galaxies).
Thus these two models serve as templates for an evolution peaking early
or late, respectively.

We perform steps 1--3 injecting pure p, He, N, Si or Fe primaries in the 
initial step and obtain then results for a mixed injection spectrum 
performing a linear combination of the final spectra.
The model parameters used for the calculations are summarized in Table~1. 
The interaction depths $\tau_0^{pp}$ 
and  $\tau_0^{p\gamma}$ given there are those for proton primaries 
at the reference energy $E_0=10^{19}$\,eV.

\begin{table*}
\begin{tabular}{lcc}
\hline\hline
source type               & only gas & photons+gas \\
power-law injection spectrum  $\alpha$ & $1.8$ &  $1.5$\\ 
maximal energy $E_{\max}/$eV & $3\times 10^{18}$ &  $6\times 10^{18}$\\ 
evolution & BL Lac & AGN \\
interaction depth $\tau_0^{pp}$ & 0.035& 0.035 \\        
\phantom{interaction depth} $\tau_0^{p\gamma}$ & 0 & 0.29 \\        
photon temperature $T/$K  & -- & 850\\
power-law index diffusion  $\delta_{Ap}$ &   $0.5$ &  $0.5$\\ 
power-law index diffusion  $\delta_{A\gamma}$ & -- &  $0.77$\\ 
\hline
\end{tabular}
\caption{Parameters used for the two cases shown in Fig.~1.}
\end{table*}

\begin{figure*}
\begin{tabular}{c}
\includegraphics[width=0.7\columnwidth,angle=270]{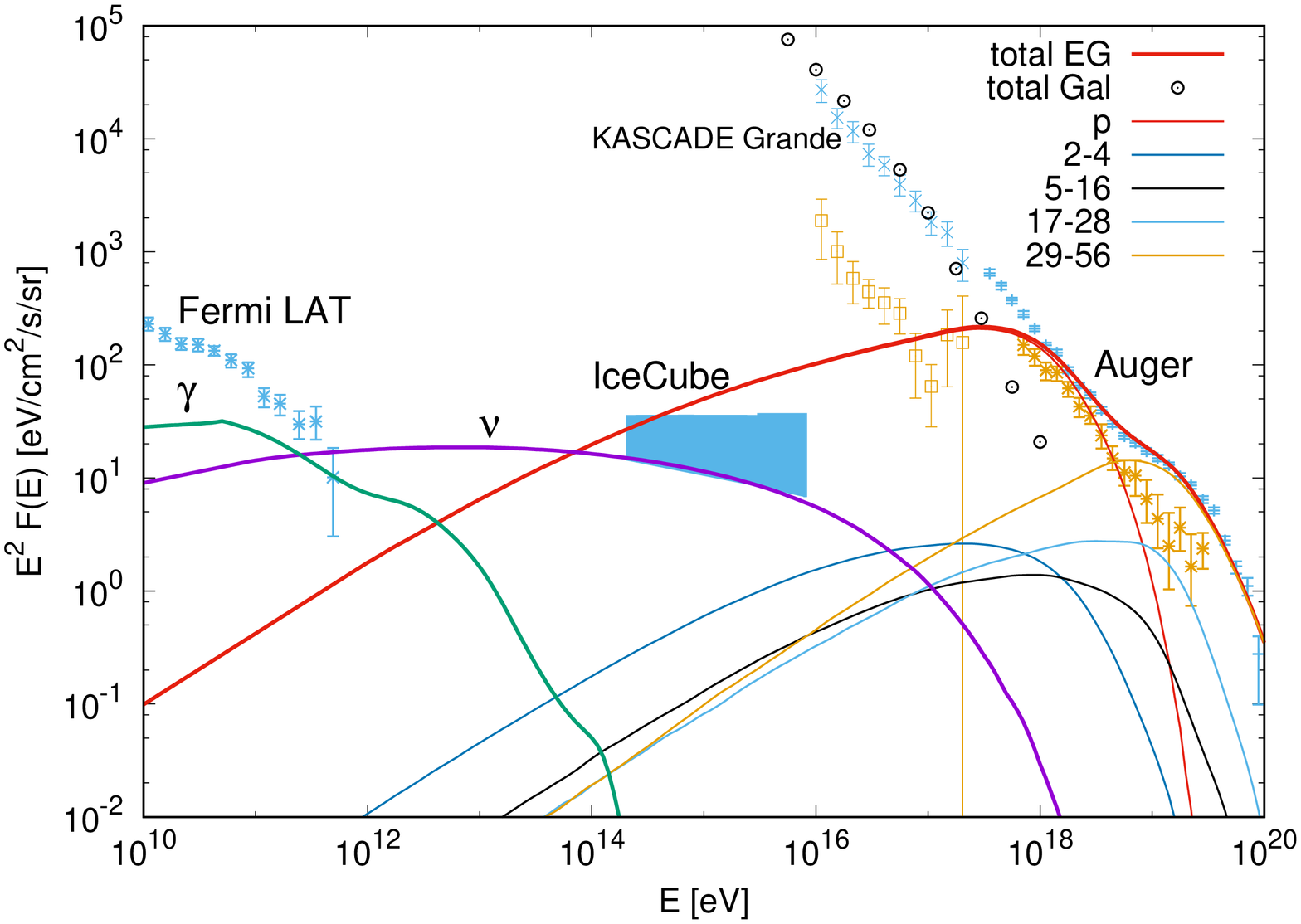}
\includegraphics[width=0.7\columnwidth,angle=270]{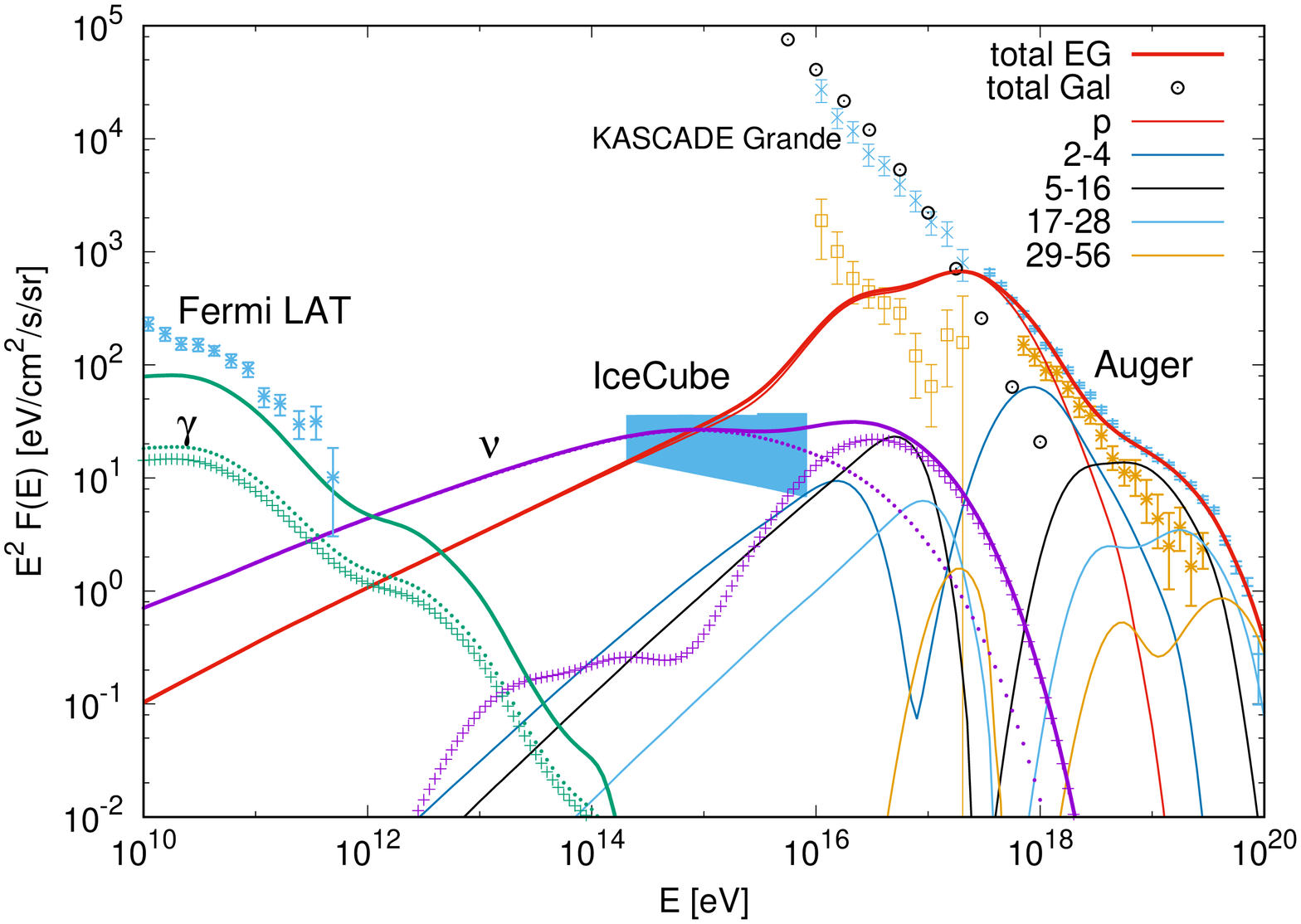}
\\
\includegraphics[width=0.7\columnwidth,angle=270]{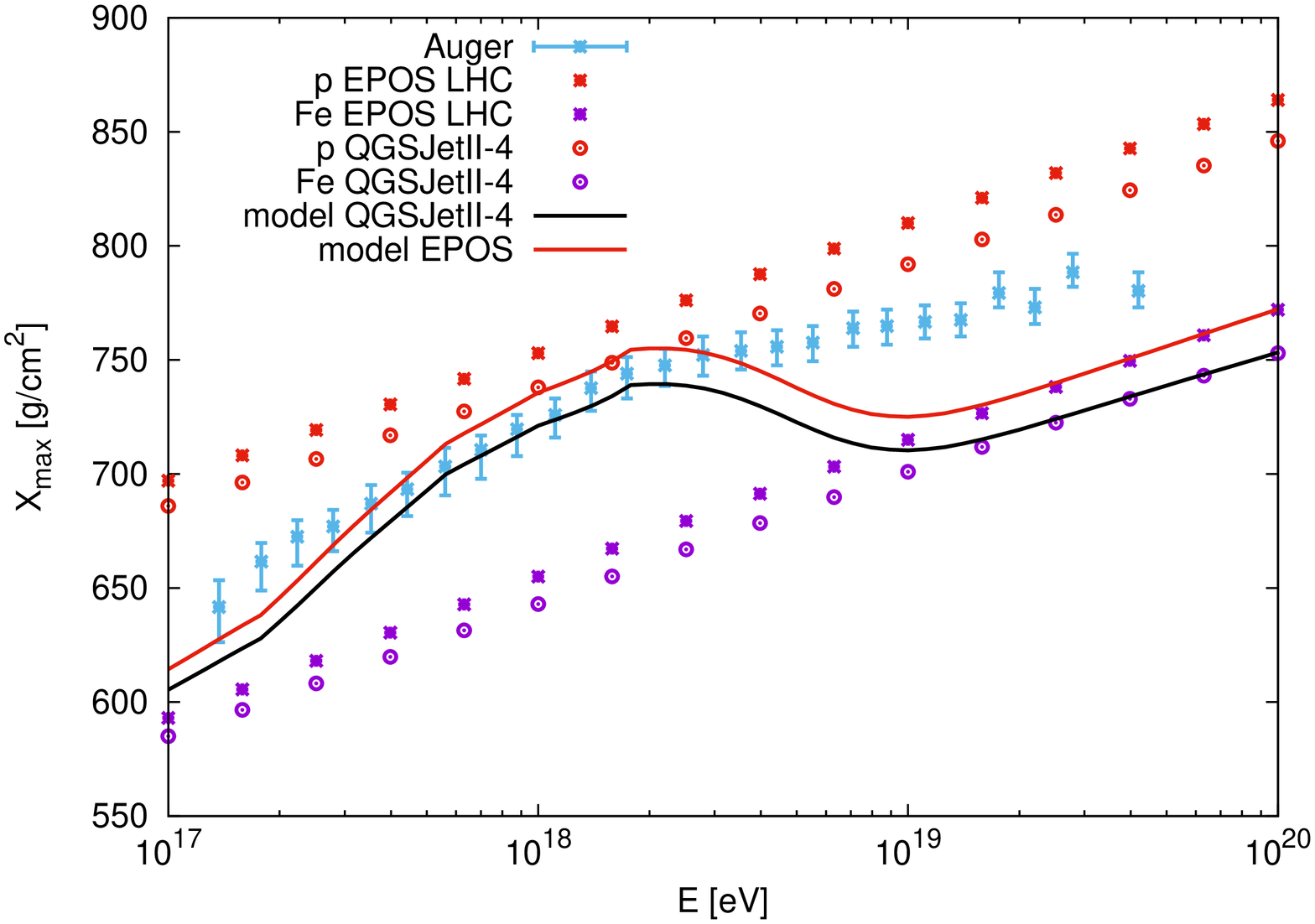}
\includegraphics[width=0.7\columnwidth,angle=270]{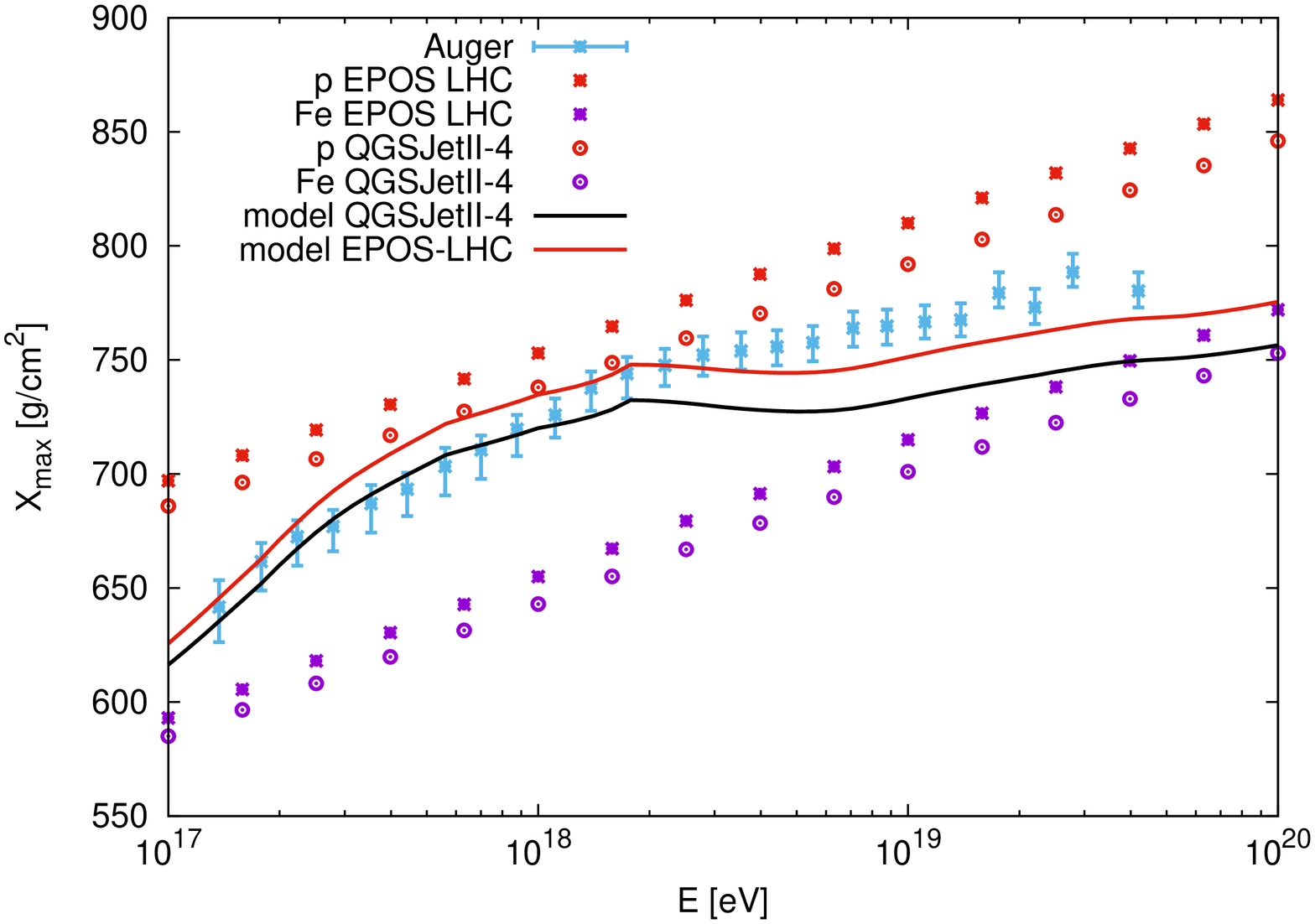}
\\
\includegraphics[width=0.7\columnwidth,angle=270]{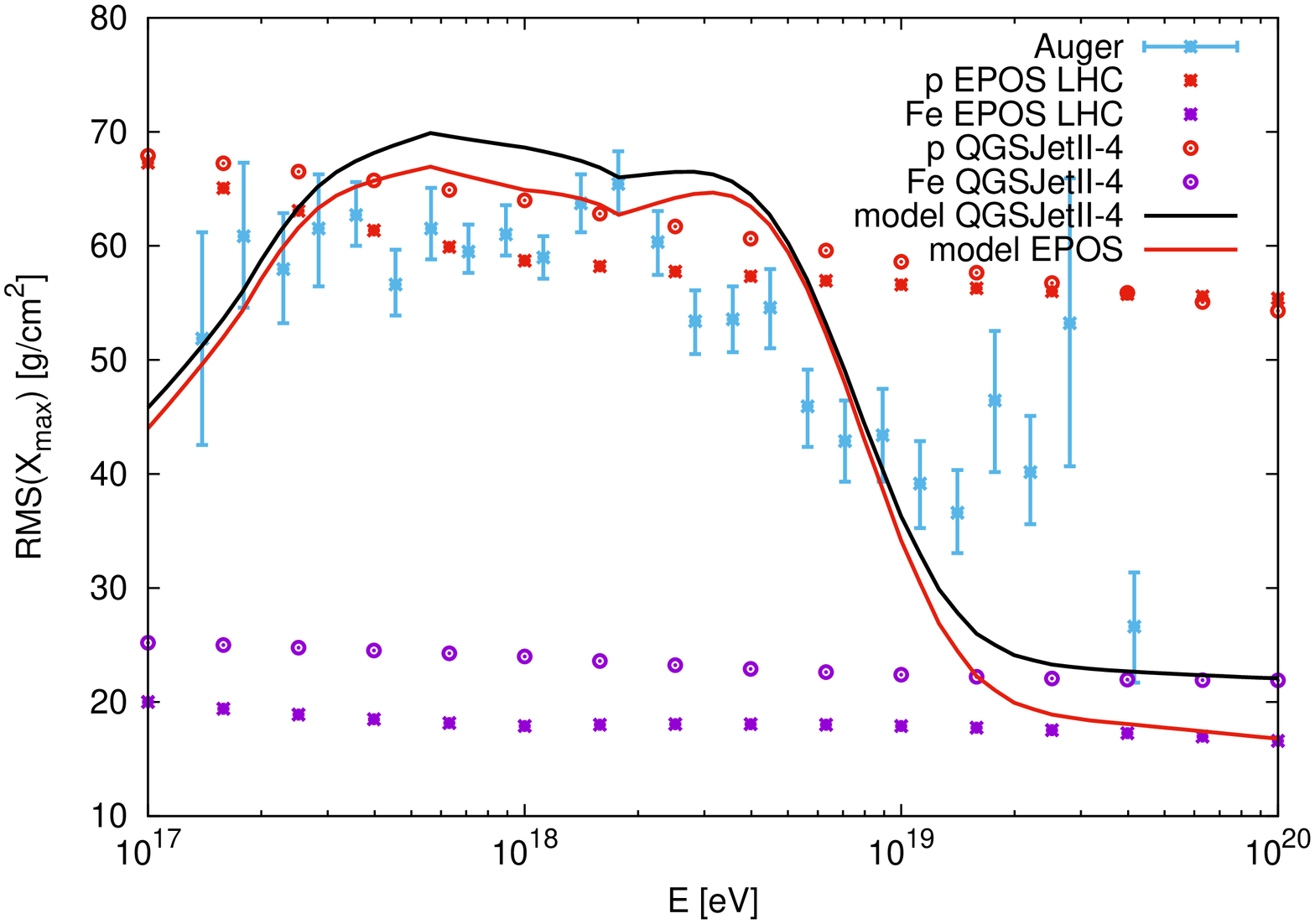}
\includegraphics[width=0.7\columnwidth,angle=270]{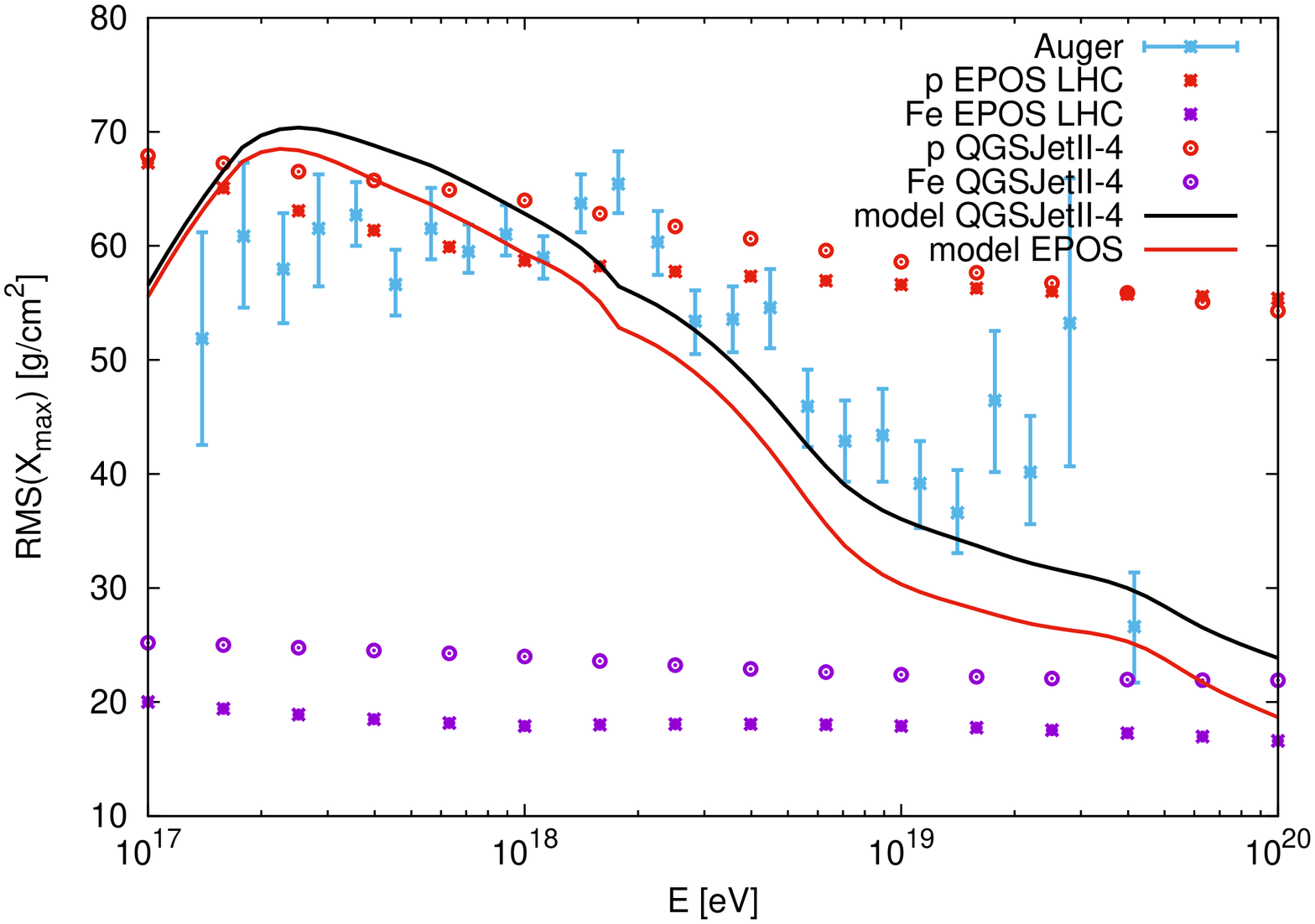}
\end{tabular}
\caption{Predictions for the diffuse flux (top) of five elemental groups
together with the proton (orange errorbars) and total flux from 
KASCADE, KASCADE-Grande  (light-blue errorbars)~\cite{dataKG} and 
Auger (dark-blue errorbars)~\cite{Aab:2013ika,Aab:2014aea}, 
the EGRB from Fermi-LAT (light-blue errorbars)~\cite{EGRB}, and 
the high-energy neutrino flux from IceCube (light-blue shaded area)~\cite{nuflux2}.
Crosses and dotted lines denote neutrinos and photons from
$A\gamma$ and $Ap$ interaction, respectively.
The middle and lower panels compare
predictions for \Xmax and \X2 using the EPOS-LHC~\cite{EPOS} and 
QGSJET-II-04~\cite{qgsjetII} models to data from Auger~\cite{XRMS}. 
Left panels for only hadronic interactions with
$\alpha=1.8$, $E_{\max}=3\times 10^{18}$\,eV and BL Lac evolution.
Right panels for both $A\gamma$ and Ap interactions with $\alpha=1.5$, 
$E_{\max}=6\times 10^{18}$\,eV, $\tau^{p\gamma}=0.29$ and AGN evolution.
The hadronic interaction depth is normalised as $\tau^{pp}_0=0.035$.
\label{fig1}}
\end{figure*}

\section{Results}

Having calculated for a specific set of source parameters the diffuse
flux of UHECR nuclei and their secondaries, we combine them with
the high-energy part of the Galactic CR fluxes predicted in the escape 
model~\cite{escape2}. We aim to reproduce the total CR 
flux~\cite{Aab:2014aea,dataKG,Aab:2013ika}, the average maximum 
depth \Xmax\ of observed  CR-induced air showers
and the width of the \Xmax\ distributions~\cite{XRMS}. Since the 
experimental and theoretical 
systematic uncertainties in both the \Xmax\ and the \X2\ 
measurements are difficult to quantify, we do not perform a standard $\chi^2$
analysis. Instead, we determine first the nuclear composition of extragalactic
CRs which fits best the observed total CR flux, since this is the most 
reliable quantity. After that we check if the composition is allowed
by the EGRB constraint and results in a sufficiently large neutrino
flux.

We consider first the case that photo-hadronic interactions are negligible.
Requiring both a small contribution to the EGRB and a large contribution
to the neutrino signal observed by IceCube restricts the allowed range of 
slopes $\alpha$ strongly, $\alpha\lsim 2.1$.  We choose $\delta_{Ap}=0.5$ as 
the energy-dependence of the escape rate 
$R_{\rm esc}=R_0 [E/(ZE_0)]^{1/2}$, 
corresponding to Kraichnan turbulence. The interaction and escape rates
have been  normalized such that $\tau^{pp}_0=0.035$
for proton primaries at $E_0=10$\,EeV. 
Since the interaction depth decreases with increasing energies
 due to the faster CR escape, the UHECR flux is dominated by 
primary nuclei. In order to reproduce the light component in the
KASKADE-Grande data, the extragalactic CR flux
has to remain  light up to $10^{18}$\,eV.
 Insisting to reproduce the ankle requires a
relatively low cutoff energy, $E_{\max}=3\times 10^{18}$\,eV,
and a small contribution of intermediate nuclei. Thus the
wish to describe the ankle including the sub-ankle region
by extragalactic CRs drives the composition towards a 
two-component model, consisting mainly of protons and iron. 
Note that in this scenario the spectra of intermediate  CNO nuclei
are cut off around the ankle energy, and hence their  contribution is
insignificant, unless  the proton flux is strongly reduced.
A reduction of the proton flux would in turn reduce the  neutrino flux
and the model will fall short of explaining the IceCube data.

The upper left panel of Fig.~\ref{fig1} shows the resulting diffuse
fluxes of CR nuclei and their secondaries. Both the ankle, which 
corresponds  in this scenario to the transition between the proton 
and iron dominated components in the extragalactic CR flux, and the 
proton (light) component observed by KASCADE-Grande are reproduced well.
The contribution to the diffuse EGRB is low except in the TeV range, while
the neutrino flux is somewhat below the level indicated by IceCube 
observations. The two lower left panels of Fig.~\ref{fig1} show
the predicted shower maximum \Xmax\ and the corresponding 
distribution width \X2, respectively. 
Since   the composition above
the ankle is heavy, the predicted \Xmax\ coincides with the one of Fe,
in contradiction  
to observations. Also the predicted width \X2\
is smaller than the observed one.

In our second scenario, we add photo-hadronic interactions close to the 
source. In contrast to hadronic interactions, nuclear scattering on IR 
photons  can give rise to relatively large interaction depths,
leading to the dominance of secondary nuclei in the ankle region.
The parameter
space of this case, however without including hadronic interactions on
gas, was extensively studied in Ref.~\cite{Unger:2015laa}. We employ
therefore simply their base case, using $T=850$\,K, $\tau^{p\gamma}=0.29$
and $\delta_{p\gamma}=0.77$ for diffusion close to the source. In contrast
to Ref.~\cite{Unger:2015laa}, we assume however a compact acceleration
region and use therefore an exponential attenuation. Moreover,
we choose as injection slope $\alpha=1.5$ and $E_{\max}=6\times 10^{18}$\,eV.
The interaction depth for hadronic interactions is kept as before,
i.e.\ normalised as $\tau_0^{pp}=0.035$.
The resulting diffuse fluxes of CR nuclei and their secondaries are shown
in the upper right panel of Fig.~\ref{fig1}. 
The spectra of intermediate nuclei show a narrow dip: The low-energy
end of this dip is determined by the threshold energy for photo-dissociation,
which is at higher energies partly filled by secondary nuclei generated
by heavier primary nuclei. 
The resulting neutrino flux matches now IceCube data, while
the contribution to the diffuse EGRB is of order 30\%.
While neutrinos from interactions on gas dominate at $\lsim 10^{16}$\,eV, 
$A\gamma$ interactions become important at higher energies.
The two lower right panels of Fig.~\ref{fig1} 
show the predicted shower maximum \Xmax\ and the corresponding 
distribution width \X2,  respectively. Accounting for the systematic 
uncertainties of the experimental data and of the hadronic interaction 
models, the two distributions are well reproduced. 
Note that the peak
in \X2 could be shifted to lower energies, reducing the 
extragalactic proton flux somewhat. This may indicate that not all
neutrons escape freely from their sources.

\section{Discussion}

Let us now comment in which astrophysical environments the considered UHECR
interactions could be realized. Interaction depths of order one for 
proton-gamma interaction arise naturally from scattering on IR
photons~\cite{PASA}. These photons may be either emitted by the dust torus 
of few parsecs extension or, as considered here, from a more compact
source region~\cite{M}. Similarly, the
dust and gas in the accretion disc surrounding the SMBH provides
a target for hadronic interactions of UHECRs. In both cases, UHECRs
have to be accelerated close to the SMBH, excluding e.g.\ acceleration
sites as radio lobes of AGN jets.
The composition of injected CRs has to be strongly enhanced towards
intermediate nuclei compared to the solar composition. This may
indicate a connection to the tidal ignition of white dwarf stars close
the the SMBH, as suggested recently in Ref.~\cite{TDE}.
Next we comment on the implication of our results for the 
EGRB. Since the AGN evolution peaks early, the spectral shape
of the photon flux is close to the ``universal shape'' obtained after many
cascade generations~\cite{Berezinsky:2016feh}, which in turn reproduces 
well the observed shape
of the EGRB. Moreover, reproducing the large neutrino flux observed 
by IceCube requires that in our model unresolved AGN contribute around 
30\% to the EGRB.

Finally, we comment on the dependence of our results on the used 
experimental data and the hadronic interaction model. Using the
data of the Telescope Array (TA) experiment~\cite{Ai} is complicated, 
since the effects of the experimental cuts and of the reconstruction 
bias have to be included. A preliminary analysis, using the method
of Ref.~\cite{Aii} to mimic these effects, showed that the
TA data favor a lighter composition than obtained here, represented by a
mixture of protons and helium nuclei. On the other hand, such a
mixture is disfavored by the PAO analysis using the correlations between 
the maximum depth $X_{\max}$ and the ground detector signal~\cite{Aiii}.
Employing in the analysis the  EPOS-LHC model would result in a somewhat
heavier UHECR composition: because of the  deeper $X_{\max}$ and smaller
 \X2  predicted by that model. However, as discussed in Refs.~\cite{Av,Avi}, 
 the  deeper $X_{\max}$ values of EPOS   are disfavored by the PAO 
 analysis of the maximal muon production depth $X^{\mu}_{\max}$~\cite{Aiv}.
In addition, Ref.~\cite{Avi} suggested
that the smaller \X2 for primary iron in EPOS results from deficiencies 
in its treatment of the fragmentation of nuclear spectators.

\section{Conclusions}

We have studied, if a single source class can explain both the flux
and the composition of extragalactic CRs including the sub-ankle region 
and the high-energy neutrinos observed by IceCube. Using only hadronic 
interactions of UHECRs with gas around their sources, we have obtained
a good fit to the CR energy spectrum,  however,
only if intermediate nuclei
are sub-dominant. Therefore the predicted maximum \Xmax\ of CR-induced air showers lies higher in the atmosphere
  than observed.  Adding photo-nuclear interactions,
  with a relatively  large interaction depth, 
we were able to reduce significantly the fraction of heavy nuclei
 in the primary   fluxes and, consequently,  to fit   satisfactorily
both the spectrum and the composition data on UHECRs.
At the same time,
the high-energy neutrino flux obtained matches IceCube measurements, 
while the contribution of unresolved UHECR sources to the EGRB is of 
order 30\%. The large interaction depth for photo-nuclear interactions
suggests that UHECRs are accelerated close to SMBHs.

\acknowledgments
S.O. acknowledges support from project OS\,481/1 of the 
Deutsche Forschungsgemeinschaft.
This research was supported in part with computational resources at NTNU 
provided by NOTUR, \url{http://www.sigma2.no}, and
at the  Theoretical Physics Division of the INR RAS with support by the RFBR, 
grant 16-29-13065 ofi-m.

\vskip0.3cm
{\it Note added:\/} While we were preparing this work for submission,
the preprint arXiv:1704.00015 appeared. The authors explain the neutrino
flux measured by IceCube by CR interactions in the galaxy clusters
surrounding UHECR sources. While the basic scheme may be considered
as a specific realisation of our first secenario, our attitude differs.
Since we require that the neutrino sources explain the observed
UHECR flux, we discard the case $\alpha\simeq 1.9-2.1$ and AGN evolution
considered in arXiv:1704.00015, because it does not fit the UHECR flux.
In contrast to our assumptions, the scenario of arXiv:1704.00015 implies 
that neutrino sources are only a subdominant class of UHECR sources.

\end{document}